\documentclass[11pt,a4paper]{article}
\usepackage[margin=1in,includefoot]{geometry}
\usepackage{amsmath,amssymb}
\usepackage[mathscr]{euscript}

\usepackage[pdfstartview=FitH,pdfpagemode=None]{hyperref}
\usepackage{cite}
\usepackage{braket}

\usepackage{authblk}

\title{Central Singularity of Three-Dimensional Kerr-de Sitter Black Holes}

\author[a]{Mauro Carlone} 
\author[a,b]{Wolfgang M\"uck\thanks{corresponding author, E-mail: mueck(at)na.infn.it}}
 
\affil[a]{Dipartimento di Fisica ``Ettore Pancini'', Universit\`a degli Studi di Napoli Federico II%
\protect\\ Via Cintia; 80126 Napoli, Italy}
\affil[b]{Istituto Nazionale di Fisica Nucleare, Sezione di Napoli %
\protect\\ Via Cintia; 80126 Napoli, Italy}
\date{\today}

\numberwithin{equation}{section}
\newcommand{\ie}{i.e.,\ }
\newcommand{\eg}{e.g.,\ }

\newcommand{\const}{\operatorname{const.}}

\newcommand{\form}[1]{\mathrm{#1}}
\newcommand{\rmd}{\form{d}}

\newcommand{\hide}[1]{}

\begin{document}
\maketitle
\abovedisplayskip=11pt plus 3pt minus 3pt
\belowdisplayskip=11pt plus 3pt minus 3pt
\abovedisplayshortskip=7pt plus 3pt minus 3pt
\belowdisplayshortskip=11pt plus 3pt minus 3pt
\jot=11pt

\begin{abstract}
{\small \noindent 
For three-dimensional Kerr-de Sitter space-time, we find the singular energy-momentum and spin tensor sources that generate the non-trivial geometry. The energy-momentum tensor is symmetric, conserved and compatible with a spinning massive point particle whose mass and angular velocity we determine. The calculation is based on the analysis of the holonomy for a closed loop around the singularity of the $SO(1,3)$ Chern-Simons gauge field appropriate for gravity in the presence of a positive cosmological constant. This holonomy is related, via the non-abelian Stokes theorem, to the singular source terms at the center. Our results may be helpful for a better understanding of the algebra of observables of a local observer in the Kerr-de Sitter space-time.
}
\end{abstract}

\newpage
\tableofcontents

\linespread{1.3}

\section{Introduction}

An interesting question about black holes that is seldom asked is ``What is the energy-momentum tensor of the black hole?'' or ``What is the source term in Einstein's equation that makes the geometry non-trivial?'' Black hole geometries typically being vacuum solutions, their sources are located in regions, which usually are excluded from space-time.\footnote{A similar statement applies to charged black holes, for which the electromagnetic field's energy-momentum tensor provides some, but not all, gravitational sources.} For the Schwarzschild black hole, for example, this question has been considered only in the mid 1990's \cite{Balasin:1993fn}, and the solution understandably requires distributional techniques \cite{Geroch:1986jjl, Steinbauer:2006qi}. 
Three-dimensional gravity provides a simpler set-up and yet is of quite a bit of interest as an exactly soluble system in terms of Chern-Simons gauge theory \cite{Achucarro:1986uwr, Witten:1988hc}. For the BTZ class of solutions \cite{Banados:1992wn}, the question about the sources was recently considered in \cite{Briceno:2024ddc}. 

Working out the details of the central singularity may also improve one's understanding of the algebra of observables of a local observer entangled with the source \cite{Chandrasekaran:2022cip, Witten:2023xze} and shed light on the construction of microscopic holographic models. In the case of low-dimensional de Sitter holography, the double-scaled SYK model (DSSYK) seems to be a very good candidate for a microscopic model describing a heavy object interacting with the de Sitter environment \cite{Susskind:2021esx, Rahman:2022jsf, Goel:2023svz, Narovlansky:2023lfz, Rahman:2023pgt, Aguilar-Gutierrez:2024nau, Verlinde:2024znh, Collier:2025lux, Narovlansky:2025tpb, Okuyama:2025hsd, Blommaert:2025eps}. In this context, solutions to three-dimensional de Sitter gravity \cite{Deser:1983nh, Park:1998qk, Banados:1998tb, Balasubramanian:2001nb} and their thermodynamic properties \cite{Park:1998qk, Ghezelbash:2001vs, Mann:2002gmd, Ghezelbash:2004af, Wang:2006eb, Sekiwa:2006qj, Wang:2006bn} have recently seen a revival of interest \cite{Tietto:2025oxn, Chakravarty:2025sbg, deBoer:2025rct, Yang:2025lme, Wang:2025jfd}.

Our work is motivated, in particular, by the work of Tietto and Verlinde \cite{Tietto:2025oxn}, who developed a microscopic quantum statistical description of the Schwarzschild-de Sitter static patch and an observer and matched the thermodynamic properties of this system with the thermodynamics of the DSSYK model. A particular feature of the DSSYK model is the appearance of two distinct temperatures, which find their counterparts as the horizon temperature and the temperature an observer at the singularity experiences.\footnote{For de Sitter black holes in more than three dimensions, the presence of more than one horizon temperature has been known for a while \cite{Gibbons:1977mu, Urano:2009xn} and is often taken as an indication that such black holes are thermally unstable.} Despite the matching of the de Sitter and DSSYK thermodynamics \cite{Tietto:2025oxn}, to our mind, a better understanding of the observer would be desirable. In particular, if one identifies the observer as part of the gravitational source at the center, the intrinsic properties of the source should play a physical role in this description. Obtaining these properties is the main goal of this paper. In order to be more general, we will include rotation, which leads us to consider Kerr-de Sitter space-time in three dimensions.

Thus, the main subject of our study is the Kerr-de Sitter metric 
\begin{equation}
\label{BH:metric}
	\rmd s^2 = -f(r) \rmd t^2 +\frac1{f(r)} \rmd r^2 + r^2 \left( \rmd \phi - \frac{J}{2r^2} \rmd t \right)^2~,
\end{equation}
with 
\begin{equation}
\label{BH:f.def}
	f(r) = \frac{(r_+^2-r^2)(r_-^2+r^2)}{r^2 \ell^2}~, 
	\qquad
	J = \frac{2r_+r_-}{\ell}~.
\end{equation}
Here, $\ell$ and $r_+$ are the de Sitter radius and the radius of the cosmological horizon, respectively, whereas $r_-$ parameterizes the angular momentum.\footnote{Formally setting $\ell\to i \ell$ and $r_-\to i r_-$ yields the BTZ solution with two horizons, $r_+$ and $r_-$ \cite{Banados:1992wn, Banados:1992gq}.} 
The metric \eqref{BH:metric} is a vacuum solution of Einstein's equation with positive cosmological constant $\Lambda=\ell^{-2}$. There is only one horizon.
Special cases of \eqref{BH:metric} are the static Schwarzschild-de Sitter space-times for $r_-=0$ and de Sitter space-time in static coordinates, given by $r_+=\ell$, $r_-=0$. 
In all cases except the de Sitter space-time there is a singularity at $r=0$, if one considers the simple topology of $\mathbb{R}^3$.\footnote{Indeed, a topological resolution of the singularity is a valid alternative, as was discussed in \cite{Banados:1992gq,Briceno:2024ddc} for the BTZ case. One can introduce a new coordinate $x=r^2$ and extend the metric \eqref{BH:metric} to $x<0$. In the rotating case, $r_-\neq0$, the extension is regular at $x=0$, but has a different topology. We thank Hideki Maeda for pointing this out to us. The fact that for an external observer a local source may be indistinguishable from a non-trivial topological structure was emphasized by Misner and Wheeler \cite{Misner:1957mt}.}

The rest of the paper is organized as follows. 
In order to accommodate a rotating source it is necessary to include torsion by using the first-order formulation of gravity. We review this formulation, together with some special and useful features in three dimensions, in section~\ref{GR}. In section~\ref{CS} we explain how the Chern-Simons formulation of de Sitter gravity with gauge group $SO(1,3)$, together with the non-abelian Stokes theorem, can be used to evaluate the relevant components of the curvature and torsion singular terms. This procedure is put into practice for the Kerr-de Sitter metric in section~\ref{KdS}, where we obtain the energy-momentum and spin tensor sources that are located at the center. In section~\ref{CH}, from these results we calculate the conserved charges, namely energy and angular momentum, as they are measured by an observer co-moving with the source. 
Finally, we conclude in section~\ref{conc}. Two sections, which are somewhat unrelated with the main line of our calculation, are included as appendices. These are a review of the Kerr-de Sitter thermodynamics (appendix~\ref{TH}) and another way of obtaining the energy-momentum tensor in the case of the conical singularity of Schwarzschild-de Sitter space-time (appendix~\ref{SdS}).

\section{First-order formulation of gravity in three dimensions}
\label{GR}

In the first order formalism, the gravitational field degrees of freedom are the dreibein, $e^a{}_\mu$, and the spin connections, $\omega^{ab}{}_\mu= \omega^{[ab]}{}_\mu$, which compose the frame and spin connection one-forms, $\form{e}^a= e^a{}_\mu \rmd x^\mu$ and $\omega^{ab}=\omega^{ab}{}_\mu \rmd x^\mu$, respectively. 
The gravitational field is described by the curvature and torsion tensors,
\begin{align}
\label{GR:curvature}
	R^{ab}{}_{\mu\nu} &= 2 \left( \partial_{[\mu} \omega^{ab}{}_{\nu]} + \omega^{a}{}_{c[\mu}\,\omega^{cb}{}_{\nu]} \right)~, \\
\label{GR:torsion}
	T^a{}_{\mu\nu} &= 2 \left( \partial_{[\mu} e^a{}_{\nu]} + \omega^{a}{}_{b[\mu}\,e^{b}{}_{\nu]} \right)~.
\end{align}
It is understood that the conversion between tensor coordinate indices (greek) and frame indices (latin) is done by contracting with $e^a{}_\mu$ or its inverse, $e^\mu{}_a$, as appropriate, and raising or lowering of indices with the appropriate metrics, $g_{\mu\nu}$ or $\eta_{ab}$. The standard coordinate representations of the curvature and torsion tensors are obtained from \eqref{GR:curvature} and \eqref{GR:torsion} via the identity
\begin{equation}
\label{GR.De}
	\nabla_\mu e^a{}_\nu \equiv \partial_\mu e^a{}_\nu -\Gamma^\lambda{}_{\mu\nu} e^a{}_\lambda +\omega^a{}_{b\mu} e^b {}_\nu=0~,
\end{equation}
where $\Gamma^\lambda{}_{\mu\nu}$ is the affine connection. 

We consider the Einstein-Hilbert action in the presence of a cosmological constant and a matter term,
\begin{equation}
\label{GR:action}
	S = \frac1{16\pi G} \int \rmd^3 x\, e\left( e^\mu{}_a e^\nu{}{}_b R^{ab}{}_{\mu\nu} -2\Lambda \right) + \int \rmd^3 x\, e\, \mathcal{L}_\mathrm{m}( \psi, \nabla_\mu \psi, e^a{}_\mu)~,
\end{equation}
where $\psi$ stands for matter fields, and the matter Lagrangian density $\mathcal{L}_\mathrm{m}$ is a scalar. 
According to the minimal coupling prescription \cite{Hehl:1976kj, Watanabe:2004nt}, the spin connections appear in $\mathcal{L}_\mathrm{m}$ only through the coupling to $\psi$, 
\begin{equation}
\label{GR.min.coupling}
	\nabla_\mu\psi = \partial_\mu \psi - \frac12 \omega^{ab}{}_\mu \Sigma_{ab} \psi~,
\end{equation}
where $\Sigma_{ab}$ are the generators of the $so(1,2)$ Lorentz algebra in the appropriate representation. We omit the indices on matter fields. 

The field equations that follow from \eqref{GR:action} are the Einstein equation\footnote{The Ricci tensor and scalar are $R_{ab}=R^c{}_{acb}$ and $R=R^a{}_a$.}
\begin{equation}
\label{GR:Eintein.eq}
	R^a{}_b -\frac12 \delta^a_b \left( R -2\Lambda\right) = 8\pi G \mathcal{T}^a{}_b~,
	\qquad 
	\mathcal{T}^a{}_b  = e^a{}_\mu \frac{\partial \mathcal{L}_\mathrm{m}}{\partial e^b{}_\mu} + \delta^a_b \mathcal{L}_\mathrm{m}~,
\end{equation}
the torsion equation,
\begin{equation}
\label{GR:torsion.eq}
	T^c{}_{ab} + 2 T^d{}_{d[a} \delta^c_{b]} = 8\pi G \mathcal{S}^c{}_{ab}~,
	\qquad 
	\mathcal{S}^c{}_{ab}  = e^c{}_\mu \frac{\partial \mathcal{L}_\mathrm{m}}{\partial \nabla_\mu \psi} \Sigma_{ab} \psi~,
\end{equation}
and the field equation for the matter field,
\begin{equation}
\label{GR:matter.eq}
	\frac{\partial \mathcal{L}_\mathrm{m}}{\partial \psi } - \nabla_\mu \frac{\partial \mathcal{L}_\mathrm{m}}{\partial \nabla_\mu \psi} + T^\nu{}_{\nu\mu} \frac{\partial \mathcal{L}_\mathrm{m}}{\partial \nabla_\mu \psi} =0~.
\end{equation}
$\mathcal{T}^a{}_b$ and $\mathcal{S}^a{}_{bc}$ are the energy-momentum and spin tensors, respectively. In the presence of torsion, the Riemann tensor is not automatically index-pair symmetric, $R_{abcd}\neq R_{cdab}$, so that neither the Ricci tensor nor the energy-momentum tensor are necessarily symmetric. 

The Bianchi identities, together with the gravitational equations of motion, imply conservation laws for the energy-momentum and spin tensors.\footnote{These can be found also by the Noether procedure \cite{Watanabe:2004nt}.}
The first Bianchi identity, 
\begin{equation}
	\label{GR:Bianchi1}
	R^\sigma{}_{[\mu\nu\rho]} = \nabla_{[\mu} T^\sigma{}_{\nu\rho]} -T^\lambda{}_{[\mu\nu} T^\sigma{}_{\rho]\lambda}~,
\end{equation}
once contracted, yields the relation
\begin{equation}
\label{GR:T.antisym}
	\mathcal{T}_{[\mu\nu]} = \frac12 \left( -\nabla_\lambda \mathcal{S}^\lambda{}_{\mu\nu} + T^\sigma{}_{\sigma \lambda} \mathcal{S}^\lambda{}_{\mu\nu} \right)~.  
\end{equation}

The second Bianchi identity,
\begin{equation}
	\label{GR:Bianchi2}
	\nabla_{[\mu} R^{\lambda\sigma}{}_{\nu\rho]} + R^{\lambda\sigma}{}_{\varphi[\mu}T^\varphi{}_{\nu\rho]}=0~,
\end{equation}
contracted twice, in general leads to
\begin{equation}
	\label{GR:Bianchi2.contracted}
	\nabla_\mu R^\mu{}_\nu - \frac12 \nabla_\nu R + R^\mu{}_\lambda T^\lambda{}_{\mu\nu} 
	-\frac12 T^\rho{}_{\mu\lambda} R^{\mu\lambda}{}_{\rho \nu} =0~.
\end{equation}
To simplify this further, one needs to take into account the properties of three-dimensional space-time. 

In three dimensions, the Weyl tensor, which is defined as the traceless part of the Riemann tensor, 
\begin{equation}
\label{GR:Weyl}
	C^{ab}{}_{cd} = R^{ab}{}_{cd} - 4 \delta^{[a}_{[c} \left( R^{b]}{}_{d]} - \frac14 \delta^{b]}_{d]} R \right)~,
\end{equation} 
vanishes identically. A simple proof of this well-known fact is a warm-up exercise for the reader. Consider the double Hodge dual of the Weyl tensor, 
\begin{equation}
\label{GR:Weyl:dual}
	(\star C \star)_i{}^j = \frac14 \epsilon_{iab} C^{ab}{}_{cd} \epsilon^{cdj}~.
\end{equation}
Because of the identity 
\begin{equation} 
\label{GR:eps.identity}
	\epsilon_{abc} \epsilon^{def} = -6 \delta_{abc}^{def}~,
\end{equation}
there is at least one trace on the right hand side of \eqref{GR:Weyl:dual}, so that $(\star C \star)_i{}^j=0$ by definition. Since the Hodge dual is invertable, one has
\begin{equation}
\label{GR:Weyl:dual.inverse}
	C^{ab}{}_{cd} = \epsilon^{abi} (\star C \star)_i{}^j \epsilon_{jcd} = 0~.
\end{equation}
Therefore, in three dimensions, \eqref{GR:Weyl} tells us that the Riemann tensor is algebraically determined by the Ricci tensor. 

After using the vanishing of \eqref{GR:Weyl} in \eqref{GR:Bianchi2.contracted} to eliminate the Riemann tensor and substituting the Einstein equation \eqref{GR:Eintein.eq}, one obtains the conservation law
\begin{equation}
	\label{GR:T.conservation}
	\nabla_\mu \mathcal{T}^\mu{}_\nu - T^\mu{}_{\mu\lambda} \mathcal{T}^\lambda{}_\nu = -\frac{\Lambda}{8\pi G} T^\mu{}_{\mu\nu}~.
\end{equation}

The vanishing of the Weyl tensor also allows to rewrite the Einstein equation as follows.  
Consider the two-forms 
\begin{equation}
\label{GR:E.def}
	\form{E}^{ab} = \form{R}^{ab} - \Lambda \,\form{e}^a\wedge \form{e}^b 
	= \frac12 R^{ab}{}_{cd}\, \form{e}^c\wedge \form{e}^d - \Lambda \,\form{e}^a\wedge \form{e}^b~.
\end{equation}
If one eliminates the Riemann tensor and uses Einstein's equation \eqref{GR:Eintein.eq}, this becomes
\begin{equation}
\label{GR:E.T}
	\form{E}^{ab} = -8\pi G \left( 2 \mathcal{T}^{[a}{}_c\, \form{e}^{b]} \wedge \form{e}^c + \mathcal{T}\, \form{e}^a\wedge \form{e}^b \right)~.
\end{equation}
Then, performing a left Hodge dual yields 
\begin{equation}
\label{GR:E.dual}
	\left(\star \form{E}\right)_i = \frac12 \epsilon_{iab} \form{E}^{ab} = -4\pi G \mathcal{T}^{j}{}_i\, \epsilon_{jab} \, \form{e}^a\wedge \form{e}^b
\end{equation}
or, in coordinate form, 
\begin{equation}
\label{GR:E.dual.coordinate}
	\left(\star \form{E}\right)_\mu = \left(\star \form{E}\right)_i e^i{}_\mu  = -4\pi G \mathcal{T}^{\nu}{}_\mu\, \epsilon_{\nu\rho\sigma} \, \rmd x^\rho \wedge \rmd x^\sigma~.
\end{equation}

Similar manipulations can be done also with the torsion tensor. Consider the double Hodge dual 
\begin{equation}
\label{GR:T.double.dual}
	(\star T \star)_{ab}{}^c = - \frac12\epsilon_{abd} T^d{}_{ef} \epsilon^{efc}~.
\end{equation} 
Using \eqref{GR:eps.identity}, this can be transformed into 
\begin{equation}
\label{GR:T.double.dual.2}
	(\star T \star)_{ab}{}^c = 3 \delta^{efc}_{abd}\, T^d{}_{ef} = T^c{}_{ab} + 2 T^d{}_{d[a} \delta^c_{b]}~,
\end{equation} 
in which we recognize the left hand side of the torsion equation \eqref{GR:torsion.eq}, \ie
\begin{equation}
\label{GR:torsion.eq.dual}
	(\star T \star)_{ab}{}^c = 8\pi G\, \mathcal{S}^c{}_{ab}~.
\end{equation}
 
In addition, \eqref{GR:T.double.dual.2} tells us that the torsion tensor is equal to its double Hodge dual if and only if it is traceless. However, this is a reasonable physical assumption, at least in our case, because it ensures the absence of products of delta function terms in the conservation laws \eqref{GR:T.antisym} and \eqref{GR:T.conservation}. Therefore, we will henceforth assume that
\begin{equation}
\label{GR:T.traceless}
	T^a{}_{ab}=0~, \qquad T^c{}_{ab} = (\star T \star)_{ab}{}^c~.
\end{equation}
Under this assumption, the torsion two-form can be rewritten as
\begin{align}
\notag
	\form{T}^a & = \frac12 T^a{}_{bc} \form{e}^b \wedge \form{e}^c = \frac12 (\star T \star)_{bc}{}^a\,  \form{e}^b \wedge \form{e}^c\\
\label{GR:T.traceless.form}
	&= - \frac14 \epsilon^{abc} \,T^\mu{}_{bc} \, \epsilon_{\mu\nu\rho} \rmd x^\nu \wedge \rmd x^\rho~.
\end{align}

\section{From Chern-Simons to the central singularity}
\label{CS}

Three-dimensional gravity can be formulated in terms of a Chern-Simons gauge theory \cite{Achucarro:1986uwr, Witten:1988hc}. In the case of de Sitter gravity ($\Lambda>0$), this theory is complex with the gauge group $SL(2,\mathbb{C})$ \cite{Witten:1989ip} and its quantization leads to an $SL(2,\mathbb{C})$ WZW conformal field theory with imaginary levels. An analysis of the asymptotic symmetries has allowed to calculate the entropy of 3d de Sitter space and asymptotically de Sitter black holes  \cite{Maldacena:1998ih, Park:1998qk, Balasubramanian:2001nb}, although the application of the Cardy formula is somewhat subtle due to the presence of complex charges.

For our purposes, however, we will not need the "dynamics" of the Chern-Simons formulation, but only its "kinematics".  
Adapting the work of \cite{Briceno:2024ddc} on BTZ space-times to the de Sitter case, we will work with the gauge group $SO(1,3)$, allowing for a real gauge algebra. Another reason for using $SO(1,3)$ instead of $SL(2,\mathbb{C})$ is that $SL(2,\mathbb{C})$ is two-to-one homomorphic to $SO(1,3)$. The same is true for their subgroups $SL(2,\mathbb{R})$ and $SO(1,2)$, the latter corresponding to frame rotations. This two-to-one homomorphism leads to overall sign ambiguities in the $SL(2,\mathbb{C})$ gauge holonomy.

Starting from the standard $so(1,3)$ generators $\mathsf{J}_{AB}$ ($A,B=0,1,2,3$), 
\begin{equation}
	\label{CS:J.def}
		(\mathsf{J}_{AB})_i{}^j = \eta_{Ai} \delta^j_B - \eta_{Bi} \delta^j_A~,
\end{equation}
which satisfy the Lorentz algebra
\begin{equation}
\label{CS:SO13}
	\left[ \mathsf{J}_{AB}, \mathsf{J}_{CD} \right] = \eta_{AD} \mathsf{J}_{BC} - \eta_{BD} \mathsf{J}_{AC} - \eta_{AC} \mathsf{J}_{BD} + \eta_{BC} \mathsf{J}_{AD}~,   
\end{equation}
we define the $so(1,3)$-valued gauge field
\begin{equation}
	\label{CS:A.def}
		\mathsf{A} = \frac12 \omega^{ab} \mathsf{J}_{ab} + \frac1{\ell} \form{e}^a \mathsf{J}_{a3}~, \qquad a,b=0,1,2~.
\end{equation}
The associated field strength,\footnote{In many contexts one uses $\mathsf{F} = d\mathsf{A} -i g \mathsf{A}\wedge \mathsf{A}$. The formal gauge coupling parameter is set to $g=i$.}
\begin{equation}
	\label{CS:F.def}
	\mathsf{F} = \rmd \mathsf{A} + \mathsf{A}\wedge \mathsf{A}~,
\end{equation}
turns out to be 
\begin{equation}
	\label{CS:F}
	\mathsf{F} = \frac12 \form{E}^{ab} \mathsf{J}_{ab} + \frac1{\ell} \form{T}^a \mathsf{J}_{a3}~,
\end{equation}
where $\form{E}^{ab}$ is the two-form defined in \eqref{GR:E.def} with $\Lambda=\ell^{-2}$.

The field strength $\mathsf{F}$ vanishes wherever the vacuum field equations \eqref{GR:Eintein.eq} and \eqref{GR:torsion.eq} are satisfied. In the case of Kerr-de Sitter space-time, this means everywhere except possibly at $r=0$. It is, then, possible to determine the singular curvature and torsion terms by evaluating the holonomy of the gauge field $\mathsf{A}$ for a (tiny) closed loop around the origin \cite{Briceno:2024ddc}. The missing link is provided by the non-abelian Stokes theorem \cite{Arefeva:1979dp, Bralic:1980ra, Broda:1995wv}, which relates the Wilson loop of a non-abelian gauge field to a surface integral involving the field strength. This is similar to, but somewhat more complicated than, the familiar abelian case. More precisely, the non-abelian Stokes theorem is
\begin{equation}
	\label{CS:Stokes}
	\mathsf{U}[\partial \xi] = \mathsf{S}[\xi]~,
\end{equation}
where $\xi(s,t)$ represents a two-dimensional, simply connected, bounded sheet, $\partial\xi$ its boundary, the left hand side is the Wilson loop along the closed curve  $\gamma=\partial \xi$ ($\mathcal{P}$ denotes path ordering),
\begin{equation}
\label{CS:Wilson.line}
	\mathsf{U}[{}_s\gamma_0] = \mathcal{P} \exp \left\{-\int\limits_0^s \rmd s' \dot\gamma^\mu(s') \mathsf{A}_\mu[\gamma(s')] \right\}~,
\end{equation}
and the right hand side is given by the sheet-dependent group element ($\mathcal{T}$ denotes $t$-ordering)
\begin{equation}
\label{CS:S.sheet}
	\mathsf{S}[\xi] = \mathcal{T} \exp \left\{-\int\limits_0^1 \rmd t \int \limits_0^1 \rmd s 
	\frac{\partial \xi^\mu}{\partial s} \frac{\partial \xi^\nu}{\partial t}  
	\mathsf{U}[{}_0\xi(t)_s] \mathsf{F}_{\mu\nu}[\xi(t,s)] \mathsf{U}[{}_s\xi(t)_0] \right\}~.
\end{equation}
For more details of the set-up see \cite{Bralic:1980ra}.

Consider a circular loop of radius $r$ around the origin of a plane parameterized by polar coordinates ($r,\phi$) and recall that a singularity may be located at $r=0$. 
Because of rotational invariance, one may always find a frame, in which $\dot{\mathsf{A}} \equiv \dot\gamma^\mu(\phi) \mathsf{A}_\mu[\gamma(\phi)]$ is independent of $\phi$. Therefore, the path ordering in \eqref{CS:Wilson.line} is irrelevant, and the Wilson loop \eqref{CS:Wilson.line} becomes a matrix exponential
\begin{equation}
\label{CS:Wilson.loop.matrix}
	\mathsf{U}[\partial \xi] = \exp\left[-2\pi \mathsf{A}_\phi(r)\right]~. 
\end{equation}
On the other hand, the integral on the right hand side of \eqref{CS:S.sheet} receives a contribution only from the origin, because the integrand vanishes anywhere else. This implies that \eqref{CS:Stokes} reduces to 
\begin{equation}
	\label{CS:Stokes.reduced}
	\exp\left[-2\pi \mathsf{A}_\phi(r)\right] = \exp\left[- \mathsf{U}(r)^{-1}\,\mathsf{f}\,\mathsf{U}(r)\right]~,
\end{equation}
with
\begin{equation}
\label{CS:f.def}
	\mathsf{f} = \int \mathsf{F}~.
\end{equation}
In \eqref{CS:Stokes.reduced}, $\mathsf{U}(r)$ is the Wilson line along a curve from an arbitrary point along the loop to the singularity at $r=0$. The integral in \eqref{CS:f.def} just picks up the weight of the delta-function term of $\mathsf{F}$ \cite{Briceno:2024ddc}. 
Therefore, by evaluating the left hand side of \eqref{CS:Stokes.reduced} we can gain information on the delta-function singularity of the curvature and torsion tensors, which stem from the source terms in the gravitational field equations. For this purpose it would be sufficient to evaluate the holonomy for an infinitesimal loop. Nevertheless, we will consider a loop with a finite radius and explicitly verify the structure of the right hand side of \eqref{CS:Stokes.reduced}. 

Considering an infinitesimal loop, \eqref{CS:Stokes.reduced} does not imply, however, that $\mathsf{f}=2\pi \mathsf{A}_\phi(0)$, because there is no unique inverse of a matrix exponential. To find a unique physical answer for $\mathsf{f}$, one must remember that de Sitter space is not only free of singularities, but also void of matter sources. Imposing, then, $\mathsf{f}=0$ for de Sitter space-time gives the desired solution. 
More precisely, writing the $4\times 4$ matrix $\mathsf{A}_\phi$ as 
\begin{equation}
	\label{CS:A.decomp}
	\mathsf{A}_\phi(r) = \mathsf{V}(r)^{-1} \mathsf{A}_{\text{diag}} \mathsf{V}(r)~,
\end{equation}
where $\mathsf{A}_{\text{diag}}$ is the diagonal matrix of the eigenvalues of $\mathsf{A}_\phi$, which must be independent of $r$ according to the structure discussed above, and $\mathsf{V}(r)$ is an appropriate non-singular transformation matrix, it is clear that
\begin{equation}
	\label{CS:exp.A.decomp0}
	\exp\left[-2\pi \mathsf{A}_\phi(r)\right] = \mathsf{V}(r)^{-1} \exp(-2\pi \mathsf{A}_{\text{diag}}) \mathsf{V}(r)
	= \mathsf{V}(r)^{-1} \exp(-2\pi \mathsf{A}_{\text{diag}} -2 \pi i \mathsf{N}) \mathsf{V}(r)~,
\end{equation}
for any $\mathsf{N}=\mathrm{diag}(n_1,n_2,n_3,n_4)$, $n_1,n_2,n_3,n_4 \in \mathbb{Z}$. Because we want to interpret the exponent as an element of $so(1,3)$, there are further conditions such as the reality of $\mathsf{V}^{-1}i\mathsf{N}\mathsf{V}$ and tracelessness, $n_1+n_2+n_3+n_4=0$. Anyway, the solution we seek obviously is 
\begin{equation}
	\label{CS:N.choice}
	\mathsf{N} = i \mathsf{A}_{\text{diag,dS}}~,
\end{equation}
which means that we can finally identify
\begin{equation}
	\label{CS:f.choice}
	\mathsf{f} = 2\pi \mathsf{V}(0)^{-1} \left(\mathsf{A}_{\text{diag}} - \mathsf{A}_{\text{diag,dS}}\right)\mathsf{V}(0)~.
\end{equation}
%

\section{Central singularity of Kerr-de Sitter space-time}
\label{KdS}

\paragraph{Holonomy calculation} 
Let us now put into practice the procedure developed in the previous section. With the help of the auxiliary functions 
\begin{equation}
\label{kds:params}
	A(r) = \frac{r_+^2-r_-^2-r^2}{\ell^2}~, \qquad B(r) = \frac{(r_+^2-r^2)(r_-^2+r^2)}{\ell^2}~,\qquad J = \frac{2r_+r_-}{\ell}~,
\end{equation}
a suitable orthonormal frame compatible with the metric \eqref{BH:metric} is given by\footnote{This frame is well-behaved for small $r$, but not, \eg at the horizon.}
\begin{equation}
\label{kds:frame}
	\form{e}^0 = \sqrt{A} \rmd t +\frac{J}{2\sqrt{A}} \rmd \phi~,\qquad
	\form{e}^1 = \frac{r}{\sqrt{B}}\rmd r~,\qquad
	\form{e}^2 = \frac{\sqrt{B}}{\sqrt{A}} \rmd \phi~.	
\end{equation} 
The non-zero components of the spin connections are 
\begin{equation}
\label{kds:spin.conn}
	\omega^{01} = -\frac{\sqrt{B}}{\ell^2\sqrt{A}} \rmd t~,\qquad
	\omega^{02} = -\frac{r J}{2\ell^2 A \sqrt{B}} \rmd r~,\qquad
	\omega^{12} = \frac{J}{2\ell^2 \sqrt{A}} \rmd t- \sqrt{A}\rmd \phi~.
\end{equation} 

For the time being, we will calculate a Wilson loop with a finite radius $r$. In addition to getting $\mathsf{f}$, this will allow us to check the structure of the $r$-dependence on the right hand side of \eqref{CS:Stokes.reduced}. Substituting \eqref{kds:frame} and \eqref{kds:spin.conn} into \eqref{CS:A.def} yields
\begin{equation}
\label{kds:A.phi}
	\mathsf{A}_\phi = -\sqrt{A} \mathsf{J}_{12} + \frac{J}{2\ell\sqrt{A}} \mathsf{J}_{03} +\frac{\sqrt{B}}{\ell\sqrt{A}} \mathsf{J}_{23}~.
\end{equation}
It is explicitly given by the matrix
\begin{equation}
\label{kds:A.phi.expl}
	\mathsf{A}_\phi(r) = \frac1{\ell}
	\begin{pmatrix}
	0 & 0 & 0 & -\frac{r_+r_-}{\sqrt{r_+^2-r_-^2-r^2}} \\
	0 & 0 & -\sqrt{r_+^2-r_-^2-r^2} & 0 \\
	0 & \sqrt{r_+^2-r_-^2-r^2} & 0 & \frac{\sqrt{(r_+^2-r^2)(r^2 + r_-^2)}}{\sqrt{r_+^2-r_-^2-r^2}} \\
	-\frac{r_+r_-}{\sqrt{r_+^2-r_-^2-r^2}} & 0 & -\frac{\sqrt{(r_+^2-r^2)(r^2 + r_-^2)}}{\sqrt{r_+^2-r_-^2-r^2}} & 0
	\end{pmatrix}~.
\end{equation}
Its eigenvalues are $\pm \frac{r_-}{\ell}, \pm \frac{ir_+}{\ell}$, which are independent of $r$, as discussed in section~\ref{CS}. Hence, $\mathsf{A}_\text{diag}$ and $\mathsf{A}_\text{diag,dS}$ read
\begin{equation}
\label{kds:A.diag}
	\mathsf{A}_\text{diag} = \frac1{\ell}
	\begin{pmatrix}
		-r_- &0 &0 &0 \\
		0 & r_- &0 &0 \\
		0 & 0& -ir_+ &0 \\
		0& 0& 0& ir_+ 
	\end{pmatrix}~, \qquad 
	\mathsf{A}_\text{diag,dS} = 
	\begin{pmatrix}
		0 &0 &0 &0 \\
		0 & 0 &0 &0 \\
		0& 0& -i &0 \\
		0& 0& 0& i 
	\end{pmatrix}~.
\end{equation}

A transformation matrix $\mathsf{V}(r)$ that diagonalizes $\mathsf{A}_\phi$, see \eqref{CS:A.decomp}, can be determined as follows. First, one solves $\mathsf{V}\mathsf{A}_\phi = \mathsf{A}_\text{diag}\mathsf{V}$, which yields a matrix $\mathsf{V}(r)$ involving four non-zero, but otherwise arbitrary complex coefficients, which may also be functions of $r$, one multiplying each row. Let us call these coefficients $f_i(r)$ ($i=0,1,2,3$), with $f_i(r)$ multiplying the $i$-th row. Clearly, the matrix inverse $\mathsf{V}(r)^{-1}$ then contains a factor of $f_i(r)^{-1}$ multiplying the $i$-th column. Because these two matrices, for $r=0$, sandwich a diagonal matrix in \eqref{CS:f.choice}, the result $\mathsf{f}$ is independent of the $f_i$. 

\paragraph{$r$-dependence of the holonomy}
Before proceeding to the calculation of $\mathsf{f}$, however, let us determine the Wilson line $\mathsf{U}(r)$ in \eqref{CS:Stokes.reduced}, which is an additional check of the non-abelian Stokes theorem. It is given by 
\begin{equation}
\label{kds:U}
	\mathsf{U}(r) = \mathsf{V}(0)^{-1}\mathsf{V}(r)~.  
\end{equation}
By the same argument as above, $\mathsf{U}(r)$ depends on the four ratios $\hat{f}_i(r)=\frac{f_i(r)}{f_i(0)}$. Imposing that $\mathsf{U}(r)\in SO(1,3)$, \ie
\begin{equation}
\label{kds:U.in.SO}
	\mathsf{U}(r) \eta \mathsf{U}(r)^T = \eta~,\qquad \det \mathsf{U}(r)=1~,
\end{equation}
provides two relations for the ratios $\hat{f}_i(r)$. Using these relations, one also finds that $\det \mathsf{V}(r)$ is independent of $r$. Still, a lot of choice remains. 
We pick the quite symmetric form
\begin{equation}
\label{kds:V.choice}
	\mathsf{V}(r) = 
	\begin{pmatrix} 
		\frac{r_+\sqrt{r_+^2-r^2}}{\sqrt{r_+^2-r_-^2-r^2}} & -\sqrt{r^2+r_-^2} &
		\frac{r_-\sqrt{r^2+r_-^2}}{\sqrt{r_+^2-r_-^2-r^2}} & \sqrt{r_+^2-r^2} \\
		-\frac{r_+\sqrt{r_+^2-r^2}}{\sqrt{r_+^2-r_-^2-r^2}} & -\sqrt{r^2+r_-^2} &
		-\frac{r_-\sqrt{r^2+r_-^2}}{\sqrt{r_+^2-r_-^2-r^2}} & \sqrt{r_+^2-r^2} \\
		-\frac{ir_-\sqrt{r^2+r_-^2}}{\sqrt{r_+^2-r_-^2-r^2}} & \sqrt{r_+^2-r^2} &
		-\frac{ir_+\sqrt{r_+^2-r^2}}{\sqrt{r_+^2-r_-^2-r^2}} & \sqrt{r^2+r_-^2} \\
		\frac{ir_-\sqrt{r^2+r_-^2}}{\sqrt{r_+^2-r_-^2-r^2}} & \sqrt{r_+^2-r^2} &
		\frac{ir_+\sqrt{r_+^2-r^2}}{\sqrt{r_+^2-r_-^2-r^2}} & \sqrt{r^2+r_-^2} 
	\end{pmatrix}~.
\end{equation}
With this expression, \eqref{kds:U} yields
\begin{multline}
\label{kds:U.choice}
	\mathsf{U}(r) = 
	\left( \begin{matrix}
		\frac{r_+^3\sqrt{r_+^2-r^2}-r_-^3\sqrt{r^2+r_-^2}}{\sqrt{r_+^2-r_-^2-r^2}(r_+^2+r_-^2)\sqrt{r_+^2-r_-^2}} & 0 \\ 
		0 & \frac{r_+\sqrt{r_+^2-r^2}+r_-\sqrt{r^2+r_-^2}}{r_+^2+r_-^2} \\
		\frac{\left(r_+\sqrt{r^2+r_-^2}-r_-\sqrt{r_+^2-r^2}\right)r_+r_-}{\sqrt{r_+^2-r_-^2-r^2}(r_+^2+r_-^2)\sqrt{r_+^2-r_-^2}} & 0 \\ 
		0 & \frac{r_-\sqrt{r_+^2-r^2}-r_+\sqrt{r^2+r_-^2}}{r_+^2+r_-^2} 
	\end{matrix} \right.\\
	\left. \begin{matrix}
			\frac{\left(r_+\sqrt{r^2+r_-^2}-r_-\sqrt{r_+^2-r^2}\right)r_+r_-}{\sqrt{r_+^2-r_-^2-r^2}(r_+^2+r_-^2)\sqrt{r_+^2-r_-^2}} & 0 \\
			0 & \frac{r_+\sqrt{r^2+r_-^2}-r_-\sqrt{r_+^2-r^2}}{r_+^2+r_-^2} \\
			\frac{r_+^3\sqrt{r_+^2-r^2}-r_-^3\sqrt{r^2+r_-^2}}{\sqrt{r_+^2-r_-^2-r^2}(r_+^2+r_-^2)\sqrt{r_+^2-r_-^2}} & 0 \\ 
			0 & \frac{r_+\sqrt{r_+^2-r^2}+r_-\sqrt{r^2+r_-^2}}{r_+^2+r_-^2} 		
		\end{matrix} \right)~.
\end{multline}
In order to confirm that this matrix describes the Wilson line from a point along the loop to the origin (the variable $r$ is the parameter of the start point), one needs to verify that
\begin{equation}
\label{kds:UdU.check}
	\left[\partial_r \mathsf{U}(r)\right] \mathsf{U}(r)^{-1} = \mathsf{A}_r = -\frac{rJ}{2\ell^2A\sqrt{B}}\mathsf{J}_{02} + \frac{r}{\ell\sqrt{B}} \mathsf{J}_{13}
\end{equation}
holds. This is indeed satisfied. Clearly, $\mathsf{U}(0)$ is the unit matrix.

\paragraph{Holonomy components}
Let us now return to the holonomy. From \eqref{CS:f.choice}, \eqref{kds:A.diag} and \eqref{kds:V.choice} one obtains\footnote{As explained above, the result for $\mathsf{f}$ is independent of the particular choice of $\mathsf{V}(0)$.}
\begin{multline}
\label{kds:f.expl}
	\mathsf{f} = \frac{2\pi}{(r_+^2+r_-^2)\sqrt{r_+^2-r_-^2}} \times \\
	\begin{pmatrix}
		0 & r_+r_-^2 & 0 & \frac{r_-[r_-^2(\ell -r_+) -r_+^3]}{\ell} \\
		r_+r_-^2 & 0 & \frac{r_-^4 + r_+^3(\ell-r_+)}{\ell} & 0 \\
		0 & - \frac{[r_-^4 + r_+^3(\ell-r_+)]}{\ell} & 0 & \frac{r_+r_- [r_-^2 - r_+(\ell-r_+)]}{\ell} \\
		\frac{r_-[r_-^2(\ell -r_+) -r_+^3]}{\ell} & 0 & - \frac{r_+r_- [r_-^2 - r_+(\ell-r_+)]}{\ell} & 0
	\end{pmatrix}~. 
\end{multline}
According to \eqref{CS:f.def} and \eqref{CS:F}, decomposing this matrix into its $so(1,3)$ components immediately gives the integrated two-forms $\int \form{E}^{ab}$ and $\int \form{T}^a$, where the integral is on a $t=\text{const}$ hypersurface and picks up the delta-function source terms.
We find the non-zero components
\begin{align}
\label{kds:E01}
	\int \form{E}^{01} &= - \frac{2\pi r_+r_-^2}{(r_+^2+r_-^2)\sqrt{r_+^2-r_-^2}} \\
\label{kds:E12}
	\int \form{E}^{12} &= \frac{2\pi [r_+^3(\ell-r_+) +r_-^4]}{\ell(r_+^2+r_-^2)\sqrt{r_+^2-r_-^2}} \\
\label{kds:T0}
	\int \form{T}^{0} &= \frac{2\pi r_-[r_+(r_+^2+r_-^2) - \ell r_-^2]}{(r_+^2+r_-^2)\sqrt{r_+^2-r_-^2}} \\
\label{kds:T2}
	\int \form{T}^{2} &= \frac{2\pi r_+r_- (r_+^2 + r_-^2 - \ell r_+)}{(r_+^2+r_-^2)\sqrt{r_+^2-r_-^2}}~. 
\end{align}

\paragraph{Energy-momentum tensor}
Let us focus on $\form{E}^{ab}$, of which we take the left Hodge dual, $(\star \form{E})_i$, and then convert it to the coordinate basis, $(\star \form{E})_\mu$. Thus, on the one hand, we get from \eqref{kds:E01} and \eqref{kds:E12}
\begin{equation}
\label{kds:Estar.coord}
	\int (\star \form{E})_\mu = \left( \frac{2\pi [r_+^3(\ell-r_+)+r_-^4]}{\ell^2(r_+^2+r_-^2)}, 0, 
	\frac{2\pi r_+ r_-[r_+(\ell-r_+) -r_-^2]}{\ell(r_+^2+r_-^2)} \right)~.
\end{equation} 

On the other hand, from \eqref{GR:E.dual.coordinate} we have
\begin{equation}
\label{kds:E.dual.coordinate}
	\int \left(\star \form{E}\right)_\mu = -4\pi G \int \mathcal{T}^{\nu}{}_\mu\, \epsilon_{\nu\rho\sigma} \, \rmd x^\rho \wedge \rmd x^\sigma 
	= -8\pi G \int \mathcal{T}^{t}{}_\mu\, r \rmd r \rmd \phi~,
\end{equation}
where we have used the fact that the integral just picks up the $r\phi$ components of the two-form, and $\epsilon_{tr\phi}=\sqrt{g^{(3)}}=r$. 
Let us assume that the energy-momentum tensor is that of a spinning point mass. Thus, consider \eqref{T.point.particle} with a trajectory 
\begin{equation}
\label{kds:worldline}
	y(\tau)= (\tau,0,\omega \tau)~,
\end{equation}
where $\omega$ is the angular velocity (with respect to coordinate time). One straightforwardly gets
\begin{equation}
\label{kds:T.explicit}
	\mathcal{T}^{\mu\nu} = \frac{m\ell}{\sqrt{r_+^2-r_-^2 +2\omega \ell r_+r_-}} \frac{\delta(r) \delta(\phi-\omega t)}{r} 
	\begin{pmatrix}
	1 & 0 & \omega \\
	0 & 0 & 0 \\
	\omega & 0 & \omega^2
	\end{pmatrix}~.
\end{equation}
After lowering the second index, this yields for the components we need
\begin{equation}
\label{kds:T.components}
	\mathcal{T}^t{}_\mu = -\frac{m}{\sqrt{r_+^2-r_-^2 +2\omega \ell r_+r_-}} \frac{\delta(r) \delta(\phi-\omega t)}{r} 
	\left( \frac{r_+^2-r_-^2+\omega \ell r_+r_-}{\ell}, 0 , r_+r_-\right)~,
\end{equation}
so that \eqref{kds:E.dual.coordinate} becomes
\begin{equation}
\label{kds:E.dual.from.T}
		\int \left(\star \form{E}\right)_\mu = \frac{8\pi G m}{\sqrt{r_+^2-r_-^2 +2\omega \ell r_+r_-}} 
			\left( \frac{r_+^2-r_-^2+\omega \ell r_+r_-}{\ell}, 0 , r_+r_-\right)~.
\end{equation}
Comparing \eqref{kds:E.dual.from.T} with \eqref{kds:Estar.coord}, we can determine the two unknown parameters $m$ and $\omega$ characterizing the source.
After a few simple steps one finds 
\begin{align}
\label{kds:mass}
	m &=  \frac1{4G\ell} \sqrt{\frac{r_+^2(\ell-r_+)^2 - r_-^4}{r_+^2+r_-^2}}~,\\
\label{kds:omega}
	\omega &= \frac{r_-}{r_+(\ell-r_+)-r_-^2}~.
\end{align}
These relations suggest that, for a physical source, the parameters must satisfy
\begin{equation}
\label{kds:rm.cond}
	r_-^2 < r_+ (\ell-r_+)~.
\end{equation}
Moreover, it is satisfying to verify that setting $r_-=0$ in \eqref{kds:mass} reproduces the result for the Schwarzschild-de Sitter case, see \eqref{sds:mass}.

\paragraph{Spin tensor} 
We can also extract the components of the spin tensor $\mathcal{S}^a{}_{bc}$ from \eqref{kds:T0}, \eqref{kds:T2}, using \eqref{GR:torsion.eq.dual} and the assumption \eqref{GR:T.traceless}. Clearly, \eqref{kds:T0} and \eqref{kds:T2} vanish for $r_-=0$, so that the torsion is identically zero in this case. This is indeed as one would expect in the absence of rotation. 
Henceforth, we consider $r_->0$, because the limits $r_-\to 0 $ and $r\to 0$ do not commute in the inverse frame $e^\mu{}_a$.

Let us assume that also the spin tensor, like the energy-momentum tensor, has support only on the trajectory of the central spinning point mass.  
Then, from \eqref{kds:T0}, \eqref{kds:T2}, and the vanishing of $\int \form{T}^1$ we immediately read off the components
\begin{equation}
\label{kds:Ta.rp}
	T^{a}{}_{r\phi} = -T^{a}{}_{\phi r} =
	\delta(r)\delta(\phi-\omega t) \frac{2\pi r_-}{(r_+^2+r_-^2)\sqrt{r_+^2-r_-^2}} 
	\begin{pmatrix}
	r_+(r_+^2 +r_-^2) -\ell r_-^2\\
	0\\
	r_+(r_+^2 + r_-^2 -\ell r_+)
	\end{pmatrix}
\end{equation}
Converting the frame index to a coordinate index yields\footnote{This involves the inverse frame $e^\mu{}_a$, which explains why the third component does not vanish in the limit $r_-\to 0$.} 
\begin{equation}
\label{kds:Tmu.rp}
	T^\mu{}_{r\phi} = -T^\mu{}_{\phi r}~ = 	
		\delta(r)\delta(\phi-\omega t) \frac{2\pi}{r_+^2+r_-^2} 
		\begin{pmatrix}
		\ell^2 r_-\\
		0\\
		r_+^2 + r_-^2 -\ell r_+
		\end{pmatrix}~.
\end{equation}

Furthermore, under the assumption \eqref{GR:T.traceless}, from \eqref{GR:T.traceless.form} we have 
\begin{equation}
\label{kds:T.dual}
	\int \form{T}^a \epsilon_{abc} = \int T^t{}_{bc} \,r \rmd r \rmd \phi~,
\end{equation}
so that from \eqref{kds:T0} and \eqref{kds:T2} we can read off
\begin{multline}
\label{kds:Tt.ab}
	T^{t}{}_{ab} = \frac{\delta(r)\delta(\phi-\omega t)}{r} 
	\frac{2\pi r_-}{(r_+^2+r_-^2)\sqrt{r_+^2-r_-^2}}\times \\
	\begin{pmatrix}
	0 & r_+(r_+^2 + r_-^2 -\ell r_+) &0 \\
	-r_+(r_+^2 + r_-^2 -\ell r_+) &0 & r_+(r_+^2 + r_-^2) -\ell r_-^2 \\
	0 & -r_+(r_+^2 + r_-^2) +\ell r_-^2 &0 
	\end{pmatrix}~.
\end{multline}
Again, converting the frame indices to coordinate indices yields
\begin{equation}
\label{kds:Tt.munu}
	T^{t}{}_{\mu\nu} = \delta(r)\delta(\phi-\omega t) 
	\frac{2\pi}{r_+^2+r_-^2}
	\begin{pmatrix}
	0 & r_+^2 + r_-^2 -\ell r_+ &0 \\
	-(r_+^2 +r_-^2 -\ell r_+) &0 & \ell^2 r_- \\
	0 & -\ell^2 r_- &0 
	\end{pmatrix}~.
\end{equation}
It is easy to check that $T^\mu{}_{\mu\nu}=0$, if the remaining components are set to zero. Indeed, they must be zero in order to ensure the absence of singular terms in contractions with the inverse frame $e^\mu{}_a$. The fact that the expressions for $T^t{}_{r\phi}$ coincide in \eqref{kds:Tmu.rp} and \eqref{kds:Tt.munu} is a nice check of consistency. For completeness, we specify the other non-zero components, besides \eqref{kds:Tt.munu}, in matrix form,
\begin{equation}
\label{kds:Tp.munu}
	T^{\phi}{}_{\mu\nu} = \delta(r)\delta(\phi-\omega t) 
	\frac{2\pi}{r_+^2+r_-^2}
	\begin{pmatrix}
	0 & 0 &0 \\
	0 &0 & r_+^2 + r_-^2 -\ell r_+ \\
	0 & -(r_+^2 +r_-^2 -\ell r_+) &0 
	\end{pmatrix}~.
\end{equation}

Finally, the spin tensor is given by simply dividing the torsion tensor by $8\pi G$. This concludes our analysis of the source terms.

\section{Conserved charges as seen by an observer}
\label{CH}

In the previous section, we have obtained the symmetric, conserved energy-momentum tensor $\mathcal{T}^{\mu\nu}$ of the source that generates the Kerr-de Sitter space-time. This allows us to directly calculate the conserved charges. For each Killing vector $\xi$, there is an associated conserved current  
\begin{equation}
\label{CH:J.def}
	J^\mu_{(\xi)}  = \mathcal{T}^{\mu\nu} \xi_\nu
\end{equation}
and a charge
\begin{equation}
\label{CH:Q.def}
	Q_{(\xi)} = \int \rmd^2 x\, \sqrt{h}\, n_\mu J^\mu_{(\xi)}~,  
\end{equation}
where the integral is over a space-like Cauchy surface with an induced metric $h_{\alpha\beta}$, and $n_\mu$ is a unit time-like normal vector on this surface.

The Kerr-de Sitter space-time has two independent Killing vectors. Let us calculate the associated charges as they appear to an observer moving along with the source at the center. The world-line of such an observer is given by \eqref{kds:worldline}, but let us rescale $\tau=\alpha\tau_o$ such that $\tau_o$ is the proper time of the observer. Thus,
\begin{equation}
\label{CH:n.mu}
	n^\mu = \dot y^\mu(\tau_o) = \alpha \left(1,0,\omega \right)~,
\end{equation}
where $\omega$ is given by \eqref{kds:omega}, and the normalization condition $n^\mu n_\mu=-1$ imposes
\begin{equation}
\label{CH:alpha.norm}
	\alpha = \frac{\ell \sqrt{r_+(\ell-r_+)-r_-^2}}{\sqrt{(r_+^2+r_-^2)[r_+(\ell-r_+)+r_-^2]}}~.
\end{equation}
One can now identify the observer's intrinsic angular velocity (with respect to proper time),
\begin{equation}
\label{CH:omega.obs}
	\omega_o = \alpha\omega = \frac{r_-\ell}{\sqrt{(r_+^2+r_-^2)[r_+^2(\ell-r_+)^2-r_-^4]}}
\end{equation}
and temperature (defined by the periodicity of Euclidean proper time)
\begin{equation}
\label{CH:T.obs}
	T_o = \alpha |T| = \frac{\sqrt{(r_+^2+r_-^2)[r_+(\ell-r_+)-r_-^2]}}{2\pi \ell r_+ \sqrt{r_+(\ell-r_+)+r_-^2}}~,
\end{equation}
where 
\begin{equation}
\label{CH:T.horizon}
	T = \frac{f'(r_+)}{4\pi} = -\frac{r_+^2 +r_-^2}{2\pi \ell^2 r_+}
\end{equation}
is the horizon temperature. 
The vector $n^\mu$ is orthogonal to a surface\footnote{This surface has the shape of a spiral around the observer's world line.} with tangent vectors 
\begin{equation}
\label{CH:tangents}
	X_r^\mu = \left(0,1,0\right)~, \qquad X_\phi^\mu = \left( -\frac{r_+r_-\ell[r_+(\ell-r_+)-r_-^2]}{r_+^3(\ell-r_+)+r_-^4},0,1 \right)~,
\end{equation}
from which one obtains the induced integral measure
\begin{equation}
\label{CH:measure}
	\rmd^2x\, \sqrt{h} = \rmd r\, \rmd \phi\, \frac{r \ell \sqrt{r_+^2+r_-^2} \sqrt{r_+^2(\ell-r_+)^2-r_-^4}}{r_+^3(\ell-r_+)+r_-^4}~.
\end{equation} 

The world-line tangent \eqref{CH:n.mu} is itself a Killing vector, and the associated charge can be interpreted as the energy measured by the observer,
\begin{align}
\notag
	E_o &= \int \rmd^2 x\, \sqrt{h}\, n_\mu n_\nu \mathcal{T}^{\mu\nu} \\
\label{CH:E.obs}
	&= \frac{\left[r_+(\ell-r_+) +r_-^2\right]^\frac32 \sqrt{r_+(\ell-r_+) -r_-^2} \sqrt{r_+^2+r_-^2}}{4G\ell [r_+^3(\ell-r_+)+r_-^4]}~.
\end{align}

The second Killing vector is associated with the rotational symmetry, 
\begin{equation}
\label{CH:Killing.rot}
	\xi^\mu = \left( \frac{\partial}{\partial \phi} \right)^\mu = \left(0,0,1 \right)~.
\end{equation}
Its associated charge is the angular momentum measured by the observer,
\begin{align}
\notag
	J_o &= \int \rmd^2 x\, \sqrt{h}\, n_\mu \xi_\nu \mathcal{T}^{\mu\nu} \\
\label{CH:J.obs}
	&= \frac{r_+r_-\left[r_+^2(\ell-r_+)^2 -r_-^4\right]}{4G\ell [r_+^3(\ell-r_+)+r_-^4]}~.
\end{align}

Interestingly, one can observe the following relation,
\begin{equation}
\label{CH:E.m.relation}
	E_o - \omega_oJ_o  = m~,
\end{equation}
where $m$ is the mass parameter \eqref{kds:mass}. This is quite suggestive, but we have not been able to find a meaningful thermodynamic law that involves the variables defined in this section, except in the special case without rotation, $r_-=0$. In that case, the observer's thermodynamic entropy agrees with the Gibbons-Hawking entropy.

\section{Conclusions}
\label{conc}

In this paper, we have analyzed the singularity of three-dimensional Kerr-de Sitter space-time and obtained the singular energy-momentum and spin tensors that generate the non-trivial geometry. This analysis was based on the holonomy of an $SO(1,3)$ Chern-Simons gauge field appropriate for gravity in de Sitter space in connection with the non-abelian Stokes theorem. For this purpose, it would have been sufficient to calculate the holonomy for an infinitesimal loop around the singularity, but we have considered a loop with a finite radius in order to check the group structure of the holonomy. A failure in this structure would have signaled a break-down of the distributional interpretation of the source terms.
The ambiguity related to the presence of multiple roots of the group (matrix) identity has been fixed by the physical condition that de Sitter space-time is source-free.  
Our main results are the explicit expressions of the symmetric energy-momentum tensor and the spin tensor. Both tensors are conserved, because the torsion is trace-free. The energy-momentum tensor is compatible with a spinning massive particle whose mass and angular velocity parameters we have determined. 
Two details of our results are worth mentioning: First, we have found a physical condition on the parameters \eqref{kds:rm.cond} that, to the best of our knowledge, has not appeared elsewhere. Second, the spin tensor, which is identically zero in the case without rotation ($r_-=0$), does not vanish in the limit $r_-\to0$. It would be interesting to interpret the coexistence of these two rotation-free solutions in terms of a phase transition at $J=0$. 

In prospect, our results may be helpful for the formulation of thermodynamic relations as they are seen by the observer, along the lines of \cite{Tietto:2025oxn}. As a first step in this direction, we have calculated the conserved charges as they should appear to an observer co-moving with the source. However, we have not been able to identify any meaningful thermodynamic laws in the presence of a non-zero angular velocity ($r_->0$), so that further investigations are needed. 
We remark that one may introduce an observer with an angular velocity that is \emph{a priori} different from that of the source, leading to an additional parameter to play with.

\section*{Acknowledgments}
The work of W.M.\ has been partially supported by the INFN research initiative STEFI.

\begin{appendix}
\section{Kerr-de Sitter Thermodynamics}
\label{TH}

The thermodynamics of Kerr-de Sitter space-time has already been discussed in \cite{Park:1998qk, Balasubramanian:2001nb, Wang:2006eb, Sekiwa:2006qj, Wang:2006bn}. There are, in fact, several versions depending on whether or not one includes the cosmological constant and/or Newton's constant as thermodynamic variables \cite{Caldarelli:1999xj, Kastor:2009wy, Kubiznak:2016qmn, Cong:2021fnf, Cong:2021jgb, Visser:2021eqk, Ahmed:2023snm, Mann:2025xrb}. Let us follow \cite{Visser:2021eqk} and include both.

For the Kerr-de Sitter metric \eqref{BH:metric}, one easily determines the following horizon temperature and angular velocity:
\begin{align}
\label{TH:temperature}
	T = \frac{f'(r_+)}{4\pi} &= -\frac{r_+^2 +r_-^2}{2\pi \ell^2 r_+}~,\\
\label{TH:Omega}
	\Omega &= \frac{r_-}{\ell r_+}~.
\end{align}
Note that the geometric definition of the temperature via the surface gravity gives rise to a negative value. The mass and angular momentum are
\cite{Park:1998qk, Balasubramanian:2001nb, Spradlin:2001pw}\footnote{We adopt the convention of \cite{Spradlin:2001pw, Tietto:2025oxn} for the mass $M$, which differs by a sign and an additive constant from the convention of \cite{Balasubramanian:2001nb}. Actually, we could multiply $\ell^2$ in the numerator of $M$ by an arbitrary number without affecting the arguments that follow. The angular momentum $J$ is, of course, just the parameter $J$ of the metric \eqref{BH:metric} with the proper physical dimensions.}
\begin{align}
\label{TH:mass}
	M &= \frac{\ell^2- r_+^2 + r_-^2}{8G\ell^2}~, \\
\label{TH:angular.momentum}
	J &= \frac{r_+r_-}{4G\ell}~, 
\end{align}
and the Bekenstein-Hawking entropy is 
\begin{equation}
\label{TH:entropy}
	S= \frac{\pi r_+}{2G}~.
\end{equation}

We have four parameters in total, $G$, $\ell$, $r_+$, and $r_-$, so that we can introduce four pairs of thermodynamic variables, namely $(S, T)$, $(J, \Omega)$, $(p, V)$, and $(N, \mu)$. In addition to the quantities listed above, we first define the central charge (a dimensionless number) 
\begin{equation}
\label{TH:N.def}
	N = \frac{\ell}{G}
\end{equation}
and the chemical potential $\mu$ by 
\begin{equation}
\label{TH:mu.def}
	\mu N = M - TS - \Omega J~. 
\end{equation}
Furthermore, we take as the volume 
\begin{equation}
\label{TH:volume}
	V = V_{n} \ell^{n}~,
\end{equation}
where $V_n$ is the volume of a unit ball in $n$ dimensions. (Reasonable values for $n$ are 1 or 2.) 
Then, if the pressure satisfies the CFT equation of state
\begin{equation}
\label{TH:pressure}
	M = nPV~,
\end{equation} 
one may easily verify that the first law holds,
\begin{equation}
\label{TH:first.law}
	\rmd M = T \rmd S + \Omega \rmd J -P \rmd V +\mu \rmd N~.
\end{equation} 
%

\section{Conical singularity of Schwarzschild-de Sitter space-time}
\label{SdS}

Let us consider Schwarzschild-de Sitter space-time and work out the properties of the matter source at $r=0$. This exercise will show that the source is compatible with a point particle and determine the value of its mass, reproducing the value given in \cite{Tietto:2025oxn}.

For $r_-=0$, the metric \eqref{BH:metric} reduces to 
\begin{equation}
\label{sds.metric}
	\rmd s^2 = - f(r) \rmd t^2 + \frac1{f(r)} \rmd r^2 + r^2 \rmd \phi^2~,
\end{equation}
where 
\begin{equation}
\label{sds.f.def}
	f(r) = \frac{r_+^2-r^2}{\ell^2}~.
\end{equation}
The coordinate domains are $t\in \mathbb{R}$, $r\in (0,\infty)$, $\phi\sim \phi+2\pi$. $r_+$ is the cosmological horizon radius, which must satisfy the physical bound $r_+\leq \ell$. For $r_+\neq\ell$, there is a conical singularity at $r=0$, which implies the existence of a point-like source term in Einstein's equation
\begin{equation}
\label{Einstein.full}
	R_{\mu\nu} -\frac12 g_{\mu\nu} R + \Lambda g_{\mu\nu} =8 \pi G \mathcal{T}_{\mu\nu}~.
\end{equation}
In order to determine the energy-momentum tensor $\mathcal{T}_{\mu\nu}$ that creates \eqref{sds.metric}, we can work out the effect of the conical singularity on the left hand side of \eqref{Einstein.full}. In the vicinity of $r=0$, the metric \eqref{sds.metric} becomes 
\begin{equation}
\label{metric.cone}
	\rmd s^2 = - \left(\frac{r_+}{\ell}\right)^2 \rmd t^2 + \left(\frac{\ell}{r_+}\right)^2 \rmd r^2 + r^2 \rmd \phi^2~,
\end{equation}
which is a cone in the spacial part. The deficit angle of the cone is given by
\begin{equation}
\label{deficit.angle}
	\psi = 2\pi \left(1-\frac{r_+}{\ell}\right)~.
\end{equation}
For physical solutions, one expects $\psi\in[0,2\pi)$, hence the restriction $0<r_+\leq \ell$. 

An easy way to find the curvature singularity at the tip of the cone is to apply the Gauss-Bonnet theorem,
\begin{equation}
\label{Gauss-Bonnet}
	\frac1{4\pi} \int_{\mathcal{M}} \rmd^2 x \sqrt{g^{(2)}} R^{(2)} +\frac1{2\pi} \int_{\partial \mathcal{M}} \rmd s \,k = \chi =1~,  
\end{equation}
where we treat the cone $\mathcal{M}$ topologically as a disc. For a finite cone of coordinate radius $r$, $\partial \mathcal{M}$ is the circle at $r=\const$, $k=\frac{r_+}{r\ell}$ the extrinsic boundary curvature, and $\chi=1$ the Euler characteristic of $\mathcal{M}$. The result for the Ricci scalar, then, is 
\begin{equation}
\label{R2}
	R^{(2)} = 2\psi \delta^{(2)}(x,0)~,
\end{equation}
where
\begin{equation}
\label{cov.delta}
	\delta^{(2)}(x,y) = \frac1{\sqrt{g^{(2)}(x)}} \delta^2(x-y) 
\end{equation}
is the covariant delta function in the two spacial dimensions. \eqref{R2} represents the singular part of the space-time Ricci scalar. Because of the identity $$R^{(2)}_{\mu\nu} = \frac12 g^{(2)}_{\mu\nu}R^{(2)}~,$$ from \eqref{Einstein.full} one finds that the only non-zero component of $\mathcal{T}_{\mu\nu}$ is 
\begin{equation}
\label{T.tt}
	\mathcal{T}_{tt} = -\frac{\psi}{8\pi G} g_{tt} \delta^{(2)}(x,0)~.
\end{equation}

Let us compare \eqref{T.tt} with the energy-momentum tensor of a point particle of mass $m$ moving along a trajectory $y^\mu(\tau)$, 
\begin{equation}
\label{T.point.particle}
	\mathcal{T}^{\mu\nu} = m \int \rmd \tau \sqrt{-h_{\tau\tau}} u^\mu u^\nu \delta^{(3)}(x,y(\tau))~,
\end{equation}
where $h_{\tau\tau} = \dot{y}^\mu \dot{y}^\nu g_{\mu\nu} = -|\dot y|^2 $ is the induced metric on the world line, $u^\mu = \frac{\dot{y}^\mu}{|\dot{y}|}$ the particle's three-velocity (satisfying  $u_\mu u^\mu =-1$), and $$\delta^{(3)}(x,y)=\frac1{\sqrt{-g}}\delta^3(x-y)$$ is the covariant delta function in three-dimensional space-time. In the form \eqref{T.point.particle}, $\mathcal{T}^{\mu\nu}$ is manifestly invariant under reparameterizations of the world line. Using center-of-mass coordinates, we are free to fix the parameterization such that $y=(\tau, 0,0)$, so that the only non-zero component of \eqref{T.point.particle} is
\begin{align}
\notag
	\mathcal{T}^{tt} &= m \int \rmd \tau \sqrt{-g_{tt}} \left( \frac1{\sqrt{-g_{tt}}}\right)^2 \frac{\delta(t-\tau) \delta^2(x)}{\sqrt{-g_{tt}} \sqrt{g^{(2)}}} \\
	&= -m g^{tt} \delta^{(2)}(x,0)~,
\end{align}
where we have used $g^{tt}=1/g_{tt}$. Finally, comparison with \eqref{T.tt} shows that the space-time \eqref{sds.metric} contains a point mass at the origin of value
\begin{equation}
\label{sds:mass}
	m = \frac{\psi}{8\pi G} = \frac1{4G} \left( 1- \frac{r_+}{\ell} \right)~.
\end{equation} 

\end{appendix}


\providecommand{\href}[2]{#2}\begingroup\raggedright\endgroup

\end{document}